\documentclass[conference]{IEEEtran}%
\pdfoutput=1
\usepackage{amsfonts}
\usepackage{amsmath}
\usepackage{amssymb}
\usepackage{graphicx}
\usepackage{braket}%
\providecommand{\U}[1]{\protect\rule{.1in}{.1in}}
\newtheorem{theorem}{Theorem}

\newenvironment{proof}[1][Proof]{\noindent\textbf{#1.} }{\ \rule{0.5em}{0.5em}}
\begin{document}

\title{Polar codes for private classical communication}
\author{\IEEEauthorblockN{Mark M.\ Wilde} \IEEEauthorblockA{\it School of Computer Science, McGill University\\Montreal, Quebec, Canada}\and \IEEEauthorblockN{Joseph M.\ Renes} \IEEEauthorblockA{\it Institut f\"ur Theoretische Physik, ETH Zurich\\Z\"urich, Switzerland}}
\maketitle

\begin{abstract}
We construct a new secret-key assisted polar coding scheme for private
classical communication over a quantum or classical wiretap channel. The security of our
scheme rests on an entropic uncertainty relation, in addition to the channel
polarization effect. Our scheme achieves the symmetric private information
rate by synthesizing \textquotedblleft amplitude\textquotedblright\ and
\textquotedblleft phase\textquotedblright\ channels from an arbitrary quantum
wiretap channel. We find that the secret-key consumption rate of the scheme
vanishes for an arbitrary degradable quantum wiretap channel. Furthermore, we
provide an additional sufficient condition for when the secret key rate vanishes,
and we suspect that satisfying this condition implies that the scheme
requires no secret key at all. Thus, this latter condition
addresses an open question from the Mahdavifar-Vardy scheme~\cite{MV10} for polar
coding over a classical wiretap channel.

\end{abstract}

Polar coding is one of the most exciting recent developments in coding and
information theory \cite{A09}. The codes are based on the \textit{channel
polarization} effect and are provably capacity-achieving with an $O\left(
N\log N\right)  $\ complexity for both encoding and decoding, where $N$ is the
number of channel uses.

Several researchers have now extended the polar coding method to a variety of
scenarios, one of which includes the secure transmission of classical data
over a wiretap channel \cite{MV10}.
The polar codes constructed in Ref.~\cite{MV10} meet a \textquotedblleft strong
security\textquotedblright\ criterion, which requires that the mutual
information between the sender's information bits and the wiretapper's channel
outputs decrease to zero as $N\rightarrow\infty$. Guha and Wilde later
exploited the ideas from Refs.~\cite{MV10,WG11} to construct polar codes for
private data transmission over a quantum wiretap channel \cite{WG11a}, and the
resulting codes achieve the symmetric private information rate whenever the
channel to the wiretapper is classical.

An important question left open from Ref.~\cite{MV10} is to determine if it is
possible for the Mahdavifar-Vardy polar coding scheme to be both reliable and
strongly secure. Ref.~\cite{WG11a} made partial progress on this question by
suggesting a simple solution: just allow for the sender and receiver to share
some secret key before communication begins. In spite of its simplicity, this
solution is undesirable from a practical perspective because it might be
difficult for the sender and receiver to establish good secret key in the
first place. Though, Proposition~22 in Ref.~\cite{MV10} proves that the rate
of secret key needed vanishes if the wiretap channel is degradable.

In this paper, we construct a new polar coding scheme for private classical
communication over a quantum wiretap channel. As classical wiretap channels
are special cases of quantum wiretap channels, all of our results here apply
to them as well. The security of our scheme has a physical basis, due to an entropic
uncertainty relation \cite{RB09}. In fact, our scheme is secure in the strongest
information-theoretic sense, against a wiretapper who has access to unbounded
quantum computational power. The new scheme offers several improvements over
the schemes from Refs.~\cite{MV10,WG11a}:

\begin{itemize}
\item The net rate of private classical communication is equal to symmetric
private information for an \textit{arbitrary} quantum channel with qubit input.

\item The secret key consumption rate vanishes for an \textit{arbitrary}
degradable quantum wiretap channel.

\item We provide an additional sufficient condition
for when the secret key
rate of our polar coding scheme vanishes (it is based on that in Ref.~\cite{RDR11}---we suspect that satisfying this
condition means the code does
not require any secret key bits at all). Since our results here apply to classical wiretap channels as
well, this result addresses the open question from Ref.~\cite{MV10}.
We show that the condition is satisfied for some example channels (including the one from Ref.~\cite{MV10})
for a wide range of interesting parameters.
\end{itemize}

We begin in the next section with some notation and definitions.
Section~\ref{sec:polar-code-scheme} overviews our private polar coding scheme and proves that it is both
reliable and secure if sufficient secret key is available.
Section~\ref{sec:vanishing-rate-degrade}\ proves that the rate of secret key
vanishes if the quantum wiretap channel is degradable, and
the same section provides an additional sufficient condition for when the
secret key rate of the polar coding scheme vanishes. Finally, we
conclude in Section~\ref{sec:conclusion}\ with a summary.

\section{Notation and Definitions}

A binary-input \textit{classical-quantum} (cq) channel $W:x\rightarrow\rho
_{x}$ prepares a quantum state $\rho_{x}$ at the output, depending on an input
classical bit $x$. Two parameters that
determine the performance of $W$ are the fidelity $F\left(
W\right)  \equiv\left\Vert \sqrt{\rho_{0}}\sqrt{\rho_{1}}\right\Vert _{1}^{2}$
and the symmetric Holevo information $I\left(  W\right)  \equiv H\left(
\left(  \rho_{0}+\rho_{1}\right)  /2\right)  -\left[  H\left(  \rho
_{0}\right)  +H\left(  \rho_{1}\right)  \right]  /2$ where $H\left(
\sigma\right)  \equiv-$Tr$\left\{  \sigma\log_{2}\sigma\right\}  $ is the von
Neumann entropy. These parameters generalize the Bhattacharya parameter and
the symmetric mutual information \cite{A09}, respectively, and are related as
$I\left(  W\right)  \approx1\Leftrightarrow F\left(  W\right)  \approx0$ and
$I\left(  W\right)  \approx0\Leftrightarrow F\left(  W\right)  \approx1$
\cite{WG11}. The channel $W$ is near perfect when $I\left(  W\right)
\approx1$ and near useless when $I\left(  W\right)  \approx0$.

\section{Private Polar Coding Scheme}

\label{sec:polar-code-scheme}

\subsection{Classical-quantum channels for complementary variables}

Consider a quantum wiretap channel $\mathcal{N}^{A^{\prime}\rightarrow BE}$
\cite{D03,1050633}\ with a qubit input system $A^{\prime}$ for the
sender Alice, an output system~$B$ for the legitimate
receiver Bob, and an output system~$E$ for the wiretapper
Eve. The channel $\mathcal{N}^{A^{\prime}\rightarrow B}$ from the sender to
the legitimate receiver arises by tracing over the wiretapper's system:
$
\mathcal{N}^{A^{\prime}\rightarrow B}\left(  \rho\right)  \equiv\text{Tr}%
_{E}\{  \mathcal{N}^{A^{\prime}\rightarrow BE}\left(  \rho\right)
\}  ,
$
where $\rho$ is some qubit input to the channel. Let $U_{\mathcal{N}%
}^{A^{\prime}\rightarrow BES_{2}}$ denote an isometric extension of this
channel \cite{W11}, such that $S_{2}$ is its environment. The system $S_{2}$ is known as a \textquotedblleft
shield\textquotedblright\ system \cite{HHHO05PRL,HHHO09}, which is not
available to the wiretapper and may or may not be available to the legitimate
receiver. The security of our scheme is in part based on the fact that this
shield system is not available to the wiretapper.

This quantum wiretap channel $\mathcal{N}^{A^{\prime}\rightarrow BE}$ captures
the case in which the wiretapper has access to all the physical degrees of
freedom that are not available to the legitimate receiver (so that $S_{2}$ is
a trivial system in this case). It also captures the more specialized case in
which the output systems $B$ and $E$ are classical systems (as in the
classical wiretap channel \cite{W75}). (See Appendix~\ref{app:cw-as-qw} for a
brief discussion of this latter point.)

Following Ref.~\cite{RB08}, we can produce a protocol for sending classical
information privately over the quantum wiretap channel $\mathcal{N}%
^{A^{\prime}\rightarrow BE}$ by considering two different complementary
channels arising from it. Both of these derived channels are cq channels that
have a classical input bit and a quantum output depending on this input bit.

The first channel that we consider has Alice prepare a quantum state $\rho
_{z}^{A^{\prime}}$ depending on the value of an input bit $z$. She then feeds
this state into the channel $\mathcal{N}^{A^{\prime}\rightarrow B}$:%
\begin{equation}
W_{A,B}:z\rightarrow\mathcal{N}^{A^{\prime}\rightarrow B}(\rho_{z}^{A^{\prime
}}). \label{eq:amplitude-channel}%
\end{equation}
We call the above channel the \textquotedblleft amplitude cq
channel,\textquotedblright\ and the notation $W_{A,B}$ indicates that
it is an amplitude ($A$) channel to Bob ($B$). The state $\rho
_{z}^{A^{\prime}}$ can generally be a mixed state, as would arise, for
instance, from using randomness at the encoder. It thus admits a
\textit{purification} $\left\vert \psi_{z}\right\rangle ^{A^{\prime}S_{1}}$,
which is some pure state on a larger tensor-product Hilbert space
$\mathcal{H}^{A^{\prime}}\otimes\mathcal{H}^{S_{1}}$ such that tracing over
the purifying system $S_{1}$ gives back the original state: $\rho
_{z}^{A^{\prime}}=\ $Tr$_{S_{1}}\{\left\vert \psi_{z}\right\rangle
\left\langle \psi_{z}\right\vert ^{A^{\prime}S_{1}}\}$. The system $S_{1}$ is
an additional shield system because it represents another system that is not
available to the wiretapper (though in this case, Alice always has access to
this shield system). By purifying all systems, we can see that the amplitude
channel in (\ref{eq:amplitude-channel}) arises by tracing over the $S_{1}%
S_{2}E$ systems of the following cq channel:%
\begin{equation}
z\rightarrow\left\vert \psi_{z}\right\rangle ^{BES_{1}S_{2}}\equiv
U_{\mathcal{N}}^{A^{\prime}\rightarrow BES_{2}}\left\vert \psi_{z}%
\right\rangle ^{A^{\prime}S_{1}}. \label{eq:purified-amplitude-channel}%
\end{equation}
The symmetric Holevo information of this amplitude channel $W_{A,B}$\ is equal
to $I\left(  W_{A,B}\right)  \equiv I\left(  Z;B\right)_\xi  $, where the
mutual information $I\left(  Z;B\right)  $ is computed with respect to%
\begin{equation}
\xi^{ZBE}\equiv\frac{1}{2}\sum_{z}\left\vert z\right\rangle \left\langle z\right\vert
^{Z}\otimes\mathcal{N}^{A^{\prime}\rightarrow BE}(\rho_{z}^{A^{\prime}}).
\label{eq:amplitude-cq-state}%
\end{equation}

The second cq channel that we consider is a phase channel with quantum side
information (QSI). Suppose now that Alice has access to an entangled state of
the following form:%
\begin{equation}
\left\vert \varphi\right\rangle ^{CA^{\prime}S_{1}}\equiv\frac{1}{\sqrt{2}%
}\sum_{z}\left\vert z\right\rangle ^{C}\left\vert \psi_{z}\right\rangle
^{A^{\prime}S_{1}}. \label{eq:phase-input}%
\end{equation}
If so, Alice could then modulate the $C$ system by applying a phase operator
$Z^{x}$ to it, depending on some bit $x$. If she is then able to transmit the
$A^{\prime}$ system through the channel $\mathcal{N}^{A^{\prime}\rightarrow
B}$ and the $C$ and $S_{1}$ systems through an identity channel, the resulting
channel to Bob is as follows:%
\begin{align}
W_{P,B} : & \, x\rightarrow\omega_{x}^{BCS_{1}S_{2}}, \label{eq:phase-channel}\\
\omega_{x}^{BCS_{1}S_{2}}\equiv & \,\text{Tr}_{E}\{  U_{\mathcal{N}%
}^{A^{\prime}\rightarrow BES_{1}}[\left(  Z^{x}\right)  ^{C}\left\vert
\varphi\right\rangle \left\langle \varphi\right\vert ^{CA^{\prime}S_{1}%
}\left(  Z^{x}\right)  ^{C}]\}  .\nonumber
\end{align}
The notation $W_{P,B}$ indicates that this is a phase ($P$) channel to Bob
($B$). By including the wiretapper $E$ as well, the channel acts as follows
(where we should subsequently trace over $E$):%
\begin{align}
x  &  \rightarrow\left(  Z^{x}\right)  ^{C}U_{\mathcal{N}}^{A^{\prime
}\rightarrow BES_{2}}\left\vert \varphi\right\rangle ^{CA^{\prime}S_{1}%
}\nonumber\\
&  =\frac{1}{\sqrt{2}}\sum_{z}\left(  -1\right)  ^{xz}\left\vert
z\right\rangle ^{C}\left\vert \psi_{z}\right\rangle ^{BES_{1}S_{2}}.
\end{align}
The symmetric Holevo information of this amplitude channel $W_{P,B}$\ is equal
to $I\left(  W_{P,B}\right)  \equiv I\left(  X;BCS_{1}S_{2}\right)_\eta $, where
the mutual information is computed using 
the following state:%
\begin{equation}
\eta^{XBCS_1S_2}\equiv\frac{1}{2}\sum_{x}\left\vert x\right\rangle \left\langle x\right\vert
^{X}\otimes\omega_{x}^{BCS_{1}S_{2}}. \label{eq:phase-cq-state}%
\end{equation}

Although it is not immediately obvious, the phase channel $W_{P,B}$ is useful in constructing polar codes for
private classical communication. Its importance stems from the fact that it is
intimately related to the amplitude channel from the sender to the wiretapper Eve,
via an uncertainty relation \cite{RB09}.
Indeed,
consider the channel from Alice's input bit $z$ to the wiretapper:%
\begin{equation}
W_{A,E}:z\rightarrow\mathcal{N}^{A^{\prime}\rightarrow E}(\rho_{z}^{A^{\prime
}}),
\end{equation}
where $\mathcal{N}^{A^{\prime}\rightarrow E}$ is the channel that results from
tracing over Bob's system after applying the wiretap channel $\mathcal{N}%
^{A^{\prime}\rightarrow BE}$. The notation $W_{A,E}$ indicates that this
channel is an amplitude ($A$) channel to Eve ($E$). The symmetric Holevo
information of this channel is equal to $I\left(  W_{A,E}\right)  \equiv
I\left(  Z;E\right)_\xi  $.

The important uncertainty relation between the channels $W_{P,B}$ and
$W_{A,E}$ is then as follows:%
\begin{equation}
I\left(  W_{P,B}\right)  +I\left(  W_{A,E}\right)  =1,
\label{eq:uncertainty-relation}%
\end{equation}
which is a special case of Lemma~2 from Ref.~\cite{RB08}. We interpret the
above uncertainty relation as \textquotedblleft\textit{if the phase channel to
Bob is nearly perfect, then the amplitude channel to Eve must be nearly
useless and vice versa}.\textquotedblright

The above uncertainty relation then enables us to construct a reliable and strongly secure polar coding scheme for 
 sending private classical data. As outlined in Section~\ref{sec:coding-scheme}\, our scheme has
the sender transmit private information bits through the synthesized channels
(in the polar coding sense) that are nearly perfect in both amplitude and
phase for Bob. The fact that these synthesized amplitude channels are nearly
perfect guarantees that Bob will be able to recover these bits reliably, and
that these synthesized phase channels are nearly perfect for Bob guarantees
that Eve will be able to recover only a negligibly small amount of information
about the bits sent through them, due to the above uncertainty relation.

Partitioning the synthesized channels according to amplitude and phase for Bob,
rather than according to amplitude for Bob and amplitude for Eve as in
Refs.~\cite{MV10,WG11a}, has the advantage that the scheme achieves the
symmetric private information rate for all quantum wiretap channels. Moreover,
we can prove that the secret key consumption rate vanishes for all degradable
quantum channels, and we can furthermore provide an additional sufficient condition
for when the secret key rate of the polar coding scheme vanishes.

\subsection{Channel Polarization}

\label{sec:polarizaton-review}Ref.~\cite{WG11} demonstrated how to construct
synthesized versions of $W$, by channel combining and splitting \cite{A09}.
The synthesized channels $W_{N}^{\left(  i\right)}$ are of the following form:%
\begin{align}
W_{N}^{\left(  i\right)  } & : \, u_{i}\rightarrow\rho_{\left(  i\right)  ,u_{i}%
}^{U_{1}^{i-1}B^{N}}, \label{eq:split-channels}\\
\rho_{\left(  i\right)  ,u_{i}}^{U_{1}^{i-1}B^{N}}  &  \equiv\sum_{u_{1}%
^{i-1}}\frac{1}{2^{i-1}}\left\vert u_{1}^{i-1}\right\rangle \left\langle
u_{1}^{i-1}\right\vert ^{U_{1}^{i-1}}\otimes\overline{\rho}_{u_{1}^{i}}%
^{B^{N}},\\
\overline{\rho}_{u_{1}^{i}}^{B^{N}}  &  \equiv\sum_{u_{i+1}^{N}}\frac
{1}{2^{N-i}}\rho_{u^{N}G_{N}}^{B^{N}},\,\,\,\,\,\,\,\,\rho_{x^{N}}^{B^{N}%
}\equiv\rho_{x_{1}}^{B_{1}}\otimes\cdots\otimes\rho_{x_{N}}^{B_{N}},\nonumber
\end{align}
where $G_{N}$ is Arikan's encoding circuit matrix built from classical CNOT\ and
permutation gates. If the channel is classical, then these states are diagonal
in the computational basis, and the above states correspond to the
distributions for the synthesized channels \cite{A09}. The interpretation of
$W_{N}^{\left(  i\right)  }$ is that it is the channel
\textquotedblleft seen\textquotedblright\ by the input $u_{i}$ if the
previous bits $u_{1}^{i-1}$ are available and if the future
bits $u_{i+1}^{N}$ are randomized. This motivates the development of a quantum
successive cancellation decoder \cite{WG11}\ that attempts to distinguish
$u_{i}=0$ from $u_{i}=1$ by adaptively exploiting the results of previous
measurements and quantum hypothesis tests for each bit decision.

The synthesized channels $W_{N}^{\left(  i\right)  }$ polarize, in the sense
that some become nearly perfect for classical data transmission while others
become nearly useless. To prove this result, one can model the channel
splitting and combining process as a random birth process \cite{A09,WG11}, and
one can demonstrate that the induced random birth processes corresponding to
the channel parameters $I(W_{N}^{\left(  i\right)  })$ and $F(W_{N}^{\left(
i\right)  })$ are martingales that converge almost surely to zero-one valued
random variables in the limit of many recursions. The following theorem
characterizes the rate with which the channel polarization effect takes hold
\cite{AT09,WG11}, and it is useful in proving statements about the performance
of polar codes for cq channels:

\begin{theorem}
\label{thm:fraction-good}Given a binary input cq channel $W$ and any
$\beta<1/2$, it holds that $\lim_{n\rightarrow\infty}\Pr_{I}\{\sqrt
{F(W_{2^{n}}^{\left(  I\right)  })}<2^{-2^{n\beta}}\}=I\left(  W\right)  $,
where $n$ indicates the level of recursion for the encoding, $W_{2^{n}%
}^{\left(  I\right)  }$ is a random variable characterizing the $I^{\text{th}%
}$ split channel, and $F(W_{2^{n}}^{\left(  I\right)  })$ is the fidelity of
that channel.
\end{theorem}

Assuming knowledge of the good and bad channels, one can then construct a
coding scheme based on the channel polarization effect, by dividing the
synthesized channels according to the following polar coding rule:%
\begin{equation}
\mathcal{G}_{N}\left(  W,\beta\right)  \equiv \big\{  i\in\left[  N\right]
:\sqrt{F(W_{N}^{\left(  i\right)  })}<2^{-N^{\beta}} \big\}  ,
\label{eq:polar-coding-rule}%
\end{equation}
and $\mathcal{B}_{N}\left(  W,\beta\right)  \equiv\left[  N\right]
\setminus\mathcal{G}_{N}\left(  W,\beta\right)  $, so that $\mathcal{G}%
_{N}\left(  W,\beta\right)  $ is the set of \textquotedblleft
good\textquotedblright\ channels and $\mathcal{B}_{N}\left(  W,\beta\right)  $
is the set of \textquotedblleft bad\textquotedblright\ channels. The sender
then transmits the information bits through the good channels and
\textquotedblleft frozen\textquotedblright\ bits through the bad ones. A
helpful assumption for error analysis is that the frozen bits are chosen
uniformly at random such that the sender and receiver both have access to
these frozen bits. Ref.~\cite{WG11} provided an explicit construction of a
quantum successive cancellation decoder that has an error probability equal to
$o(2^{-N^{\beta}})$---let $\{\Lambda_{u_{\mathcal{A}}}^{\left(  u_{\mathcal{A}%
^{c}}\right)  }\}$ denote the corresponding decoding positive operator-valued
measure (POVM) \cite{W11}, with $u_{\mathcal{A}}$ the information bits and
$u_{\mathcal{A}^{c}}$ the frozen bits.

For our polar coding scheme for private communication, we can imagine that the
classical bits fed into the encoder are encoded into the classical states
$\left\vert 0\right\rangle $ and $\left\vert 1\right\rangle $ and that the
encoder is a coherent, quantum version of Arikan's encoder \cite{A09}, meaning
that the gates are quantum CNOTs and permutations (i.e., the same encoder as
in Refs.~\cite{RDR11,WG11a}). This perspective is helpful in our subsequent
analysis of the phase channels, though it need not be the case in
practice---the scheme will work perfectly well if the bits are just classical
bits and the encoder is Arikan's classical encoder.\textbf{ }When sending
amplitude-basis classical information through the encoder and channels, the
effect is to induce synthesized channels $W_{A,N}^{\left(  i\right)  }$ as
described above. Theorem~\ref{thm:fraction-good}\ states that the fraction of
amplitude-good channels (according to the criterion in
(\ref{eq:polar-coding-rule})) is equal to $I\left(  Z;B\right)_\xi$.

We now consider how the encoder induces synthesized channels for the
phase-basis classical information. It is important to keep in mind for our
development in this paper that these channels are merely virtual---we just
consider them in order to relate back to the amplitude channels to Eve via an
uncertainty relation (this uncertainty relation will be a slightly modified
version of that in (\ref{eq:uncertainty-relation})). The benefit of this
approach is that we can broaden the results of Refs.~\cite{MV10,WG11a}\ and
make sharper statements about the code's secret key consumption.

Proceeding similarly to Ref.~\cite{RDR11,WR12}, the same encoding operation leads to channel
polarization for the phase channel $W_{P,B}$ as well. Suppose Alice modulates
her halves of the entangled pairs as in the definition of $W_P$, but then inputs them to the
coherent encoder before sending them via the channel to Bob. The result is
\[
\frac{1}{\sqrt{2^{N}}}\sum_{z^{N}\in\{0,1\}^{N}}(-1)^{x^{N}\cdot z^{N}}%
|\psi_{z^{N}G_{N}}\rangle^{B^{N}E^{N}S_{1}^{N}S_{2}^{N}}\left\vert
z^{N}\right\rangle ^{C^{N}},
\]
whose $B^{N}S_{1}^{N}S_{2}^{N}C^{N}$ marginal state is simply $U_{\mathcal{E}%
}^{C^{N}}\omega_{x^{N}G_{N}^{T}}^{B^{N}S_{1}^{N}S_{2}^{N}C^{N}}U_{\mathcal{E}%
}^{\dagger C^{N}}$, where $U_{\mathcal{E}}$ denotes the polar encoder. Here we
have used the fact that the matrix corresponding to $G_{N}$ is invertible.
Thus, the coherent encoder also induces synthesized channels $W_{P,N}^{(i)}$
using the encoding matrix $G_{N}^{T}$ instead of $G_{N}$, modulo the
additional $U_{\mathcal{E}}$ acting on $C^{N}$. 
Theorem~\ref{thm:fraction-good}\ states that the fraction of phase-good
channels to Bob (according to the criterion in (\ref{eq:polar-coding-rule}))
is approximately equal to $I\left(  X;BCS_{1}S_{2}\right)_\eta  $. 

Note that the classical side
information for the $W_{P,N}^{\left(  i\right)  }$ is different from that in
(\ref{eq:split-channels}) because the direction of all CNOT\ gates is flipped
due to the transpose of $G_{N}$ when acting on phase variables. This means 
means that the $i^{\text{th}}$ synthesized
phase channel $W_{P,N}^{\left(  i\right)  }$ is such that all of the
\textit{future} bits $x_{N}\cdots x_{i+1}$ are available to help in decoding
bit $x_{i}$ while all of the \textit{previous} bits $x_{i-1}\cdots x_{1}$ are
randomized. (This is the same as described in Ref.~\cite{RDR11}\ for Pauli channels.)

A clear advantage of the current approach over the
previous construction from Ref.~\cite{WG11a}\ is that
Theorem~\ref{thm:fraction-good} directly applies to the phase-good channels
with the \textquotedblleft goodness criterion\textquotedblright\ given by
(\ref{eq:polar-coding-rule}). Ref.~\cite{WG11a} considered the amplitude
channels to Eve (rather than the phase-good channels to Bob), and it seemed
only possible to prove polarization results for quantum wiretap channels in
which the amplitude channel to Eve is classical. Our approach here overcomes
this difficulty by appealing to Theorem~\ref{thm:fraction-good} directly for
polarization and later relating the phase channels to Bob and the amplitude
channels to Eve via an uncertainty relation (similar to the approach
from Ref.~\cite{WR12}).

\subsection{Private Polar Coding Scheme}

\label{sec:coding-scheme}We can now specify our polar coding scheme. Divide
the synthesized cq amplitude channels $W_{A,B,N}^{\left(  i\right)  }$ into
sets $\mathcal{G}_{N}\left(  W_{A,B},\beta\right)  $ and $\mathcal{B}%
_{N}\left(  W_{A,B},\beta\right)  $ according to (\ref{eq:polar-coding-rule}),
and similarly, divide the synthesized cq phase channels $W_{P,B,N}^{\left(
i\right)  }$ into sets $\mathcal{G}_{N}\left(  W_{P,B},\beta\right)  $ and
$\mathcal{B}_{N}\left(  W_{P,B},\beta\right)  $, where $\beta<1/2$. The
synthesized channels correspond to particular inputs to the encoding
operation, and thus the set of all inputs divides into four groups:\ those
that are good for both the amplitude and phase variable, those that are good
for amplitude and bad for phase, bad for amplitude and good for phase, and
those that are bad for both variables. We denote these
channels as follows:%
\begin{align*}
\mathcal{A}  &  \equiv\mathcal{G}_{N}\left(  W_{A,B},\beta\right)
\cap\mathcal{G}_{N}\left(  W_{P,B},\beta\right)  ,\\
\mathcal{X}  &  \equiv\mathcal{G}_{N}\left(  W_{A,B},\beta\right)
\cap\mathcal{B}_{N}\left(  W_{P,B},\beta\right)  ,\\
\mathcal{Z}  &  \equiv\mathcal{B}_{N}\left(  W_{A,B},\beta\right)
\cap\mathcal{G}_{N}\left(  W_{P,B},\beta\right)  ,\\
\mathcal{B}  &  \equiv\mathcal{B}_{N}\left(  W_{A,B},\beta\right)
\cap\mathcal{B}_{N}\left(  W_{P,B},\beta\right)  .
\end{align*}

Our polar coding scheme for private classical communication has the sender
transmit information bits through the inputs in $\mathcal{A}$, random bits
through the inputs in $\mathcal{X}$, frozen bits through the inputs in
$\mathcal{Z}$, and halves of secret key bits through the inputs in
$\mathcal{B}$. It is straightforward to prove that the net rate of private
classical communication $\left(  \left\vert \mathcal{A}\right\vert -\left\vert
\mathcal{B}\right\vert \right)  /N$ is equal to the symmetric private
information $I\left(  Z;B\right)_\xi  -I\left(  Z;E\right)_\xi  $ by observing that
the fraction of amplitude-good channels is $I\left(  Z;B\right)_\xi $, the
fraction of phase-good channels is $I\left(  X;BCS_{1}S_{2}\right)_\eta $, and
exploiting the uncertainty relation $I\left(  X;BCS_{1}S_{2}\right)_\eta
=1-I\left(  Z;E\right)_\xi  $ from (\ref{eq:uncertainty-relation}). A detailed
proof is similar to the proof given in Appendix~A of Ref.~\cite{WR12}.

We should stress that our consideration of the phase channels in this paper is
only necessary in order to compute the index sets $\mathcal{A}$, $\mathcal{X}%
$, $\mathcal{Z}$, and $\mathcal{B}$. The decoder in the next section does not
make explicit use of these phase channels---they only arise in our security
analysis, where we appeal to an entropic uncertainty relation in order to
guarantee security of the scheme. This is in contrast to our polar
coding scheme for sending {\it quantum} information \cite{WR12}, in which the
decoder makes explicit use of the phase channels.

\subsection{Reliability and Security Analysis}

First, it is straightforward to prove that the code has good reliability, by
appealing to the results from Ref.~\cite{WG11}. That is, there exists a
POVM\ $\{\Lambda_{u_{\mathcal{A}},u_{\mathcal{X}}}^{\left(  u_{\mathcal{B}%
}\right)  }\}$ such that%
\[
\Pr\{\widehat{U}_{\mathcal{C}}\neq U_{\mathcal{C}}\}\leq\sqrt{2\sum
_{i\in\mathcal{C}}\sqrt{F(W_{A,B,N}^{\left(  i\right)  })}}=o\left(
2^{-\frac{1}{2}N^{\beta}}\right)  .
\]
where $\mathcal{C\equiv A}\cup\mathcal{X}$. This POVM\ is the quantum
successive cancellation decoder established in Ref.~\cite{WG11}. The quantum
successive cancellation decoder operates exactly as before, but it needs to
decode both the information bits in $\mathcal{A}$ and the randomized bits in
$\mathcal{X}$. It also exploits the frozen bits in $\mathcal{Z}$ and the
secret key bits in $\mathcal{B}$ to help with decoding. This decoder has an
efficient implementation if the channel to Bob is classical \cite{A09}. This
is the case for the amplitude damping channel and any Pauli channel, for example.

We now prove that strong security, in the sense of Ref.~\cite{MV10}, holds for
our polar coding scheme.

\begin{theorem}
For sufficiently large $N$, the private polar coding scheme given above satisfies the following strong
security criterion: $I\left(
U_{\mathcal{A}};E^{N}\right)  =o(2^{-\frac{1}{2}N^{\beta}})$.
\end{theorem}

\begin{proof}
Consider that%
\begin{align*}
I\left(  U_{\mathcal{A}};E^{N}\right)   &  =\sum_{i\in\mathcal{A}}I\left(
U_{i};E^{N}|U_{\mathcal{A}_{i}^{-}}\right)  =\sum_{i\in\mathcal{A}}I\left(
U_{i};E^{N}U_{\mathcal{A}_{i}^{-}}\right) \\
&  \leq\sum_{i\in\mathcal{A}}I\left(  U_{i};E^{N}U_{1}^{i-1}\right)
=\sum_{i\in\mathcal{A}}I(W_{A,E,N}^{\left(  i\right)  })
\end{align*}
The first equality is from the chain rule for quantum mutual information and
by defining $\mathcal{A}_{i}^{-}$ to be the indices in $\mathcal{A}$ preceding
$i$. The second equality follows from the assumption that the bits in
$U_{\mathcal{A}_{i}^{-}}$ are chosen uniformly at random. The first inequality
is from quantum data processing. The third equality is from the definition of
the synthesized channels $W_{A,E,N}^{\left(  i\right)  }$. Continuing, we have%
\begin{align*}
&  \leq\sum_{i\in\mathcal{A}}\sqrt{1-F(W_{A,E,N}^{\left(  i\right)  })}%
\leq\sum_{i\in\mathcal{A}}\sqrt{1-(1-2F(W_{P,B,N}^{\left(  i\right)  }%
)^{\frac{1}{2}})^{2}}\\
&  \leq\sum_{i\in\mathcal{A}}\sqrt{4\sqrt{F(W_{P,B,N}^{\left(  i\right)  })}%
}\leq2\sum_{i\in\mathcal{A}}\sqrt{2^{-N^{\beta}}}=o\left(  2^{-\frac{1}%
{2}N^{\beta}}\right)  .
\end{align*}
The first inequality is from Proposition~1 in Ref.~\cite{WG11}. The second
inequality follows from a fidelity uncertainty relation%
\begin{equation}
\sqrt{F(W_{A,E,N}^{\left(  i\right)  })}+2\sqrt{F(W_{P,B,N}^{\left(  i\right)
})}\geq1, \label{eq:fidelity-uncertainty}%
\end{equation}
proved in Appendix~C of Ref.~\cite{WR12}. The fourth inequality follows from
the definition of the set $\mathcal{A}$.
\end{proof}

\section{Vanishing Secret-Key Rate}

\label{sec:vanishing-rate-degrade}The rate of secret key required by our polar
coding scheme vanishes whenever the quantum wiretap channel $\mathcal{N}%
^{A^{\prime}\rightarrow BE}$ is \textit{degradable}, meaning that there exists
some degrading map $\mathcal{D}^{B\rightarrow E}$\ that allows the legitimate
receiver to simulate the output of the wiretapper: $\mathcal{D}^{B\rightarrow
E}\circ\mathcal{N}^{A^{\prime}\rightarrow B}=\mathcal{N}^{A^{\prime
}\rightarrow E}$. This condition holds for many channels of interest such as
the amplitude damping channel and the dephasing channel.

The argument proceeds similarly to the argument in Ref.~\cite{WR12}. The
argument there demonstrates that the entanglement consumption rate of a
quantum polar code vanishes for degradable channels (this argument in turn is
similar to the original argument in Ref.~\cite{MV10}). We merely highlight the
argument and point the interested reader to Appendix C of Ref.~\cite{WR12} for
the details. From the fidelity uncertainty relation in
(\ref{eq:fidelity-uncertainty}), we know that the phase-good channels to Bob
should be amplitude-\textquotedblleft very bad\textquotedblright\ to Eve, in
the sense that%
\[
\sqrt{F(W_{P,B,N}^{\left(  i\right)  })}<2^{-N^{\beta}}\Longrightarrow
\sqrt{F(W_{A,E,N}^{\left(  i\right)  })}>1-2\cdot2^{-N^{\beta}}.
\]
From degradability, we also know that the doubly-bad channels in $B$ are
amplitude-bad channels to Eve (if they are bad for Bob, then they are worse
for Eve.). These observations imply that the phase-good channels to Bob, the
doubly-bad channels to Bob, and the amplitude-good channels to Eve are
disjoint sets. From Theorem~\ref{thm:fraction-good}, the sum rate of the
phase-good channels to Bob and the amplitude-good channels to Eve is equal to
$I(W_{P,B})+I(W_{A,E})=1$ as $N\rightarrow \infty$. Thus, the rate of the doubly-bad set is zero in the same limit.


The theorem below provides another sufficient condition for
when our private polar code has a vanishing secret key rate. The argument
proceeds along the lines given in Section~7.1 of Ref.~\cite{RDR11}, with the
fidelity replacing the Bhattacharya parameter. We provide a proof in
Appendix~\ref{app:no-secret-key-proof}\ for completeness.

\begin{theorem}
\label{thm:suff-cond}If the following inequality holds, then the private polar
coding scheme has a vanishing secret key rate:%
\[
\sqrt{F(W_{A,B})}+\sqrt{F(W_{P,B})}<1.
\]
$F(W_{A,B})$ and $F(W_{P,B})$ are the fidelities of the amplitude channel in
(\ref{eq:amplitude-channel}) and the phase channel in (\ref{eq:phase-channel}%
), respectively.
\end{theorem}

It is our suspicion
that channels satisfying the above condition do not require any secret key at all, but we have
not been able to prove it (we discuss this point further in Appendix~\ref{app:no-secret-key-proof}). 
We note that the above theorem provides a similar sufficient condition for the
quantum polar codes from Ref.~\cite{WR12}\ to determine if the codes there have a vanishing rate
of entanglement consumption.

In Appendix~\ref{app:example-channels}, we compute $\sqrt{F(W_{A,B})}+\sqrt{F(W_{P,B})}$ for
several example channels, including the binary symmetric wiretap channel from Ref.~\cite{MV10},
the erasure wiretap channel, and the amplitude damping channel. We find that the secret key rate
vanishes for a wide range of interesting parameters.

\section{Conclusion}

\label{sec:conclusion}Building on the
general approach from Ref.~\cite{RB08}, we have constructed a polar coding scheme for private
classical communication over a quantum wiretap channel which achieves the symmetric private information rate. By considering the associated classical amplitude and phase channels 
of the wiretap channel, we are able to demonstrate that the scheme is both reliable and strongly secure. Indeed, the reliability of non-private polar coding is sufficient, as the strong security of the private protocol follows from the reliability of the phase channel coding. Additionally, we have shown that the secret-key
consumption rate vanishes for all degradable quantum wiretap channels or channels satisfying a simple fidelity criterion. 
As classical wiretap channels are a special form of quantum wiretap channels, all of our results apply to them as well. It would be interesting to find an argument ensuring reliability and strong security of the scheme which relies only on a classical description of the wiretap channel. Further important open questions include whether 
 there is an efficient implementation of the quantum successive cancellation decoder used here,
if there is a fast algorithm for determining the good and bad channels, and if
the condition in Theorem~\ref{thm:suff-cond} implies that the codes require no secret key at all.

We thank Frederic Dupuis for useful feedback on Theorem~\ref{thm:suff-cond}.
MMW\ acknowledges support from the Centre de Recherches Math\'{e}matiques, and
JMR\ acknowledges support from the Swiss National Science Foundation and the European Research Council.
\bibliographystyle{IEEEtran}
\bibliography{Ref}

\pagebreak

\onecolumn

\appendices

\section{Classical Wiretap Channels as Quantum Wiretap Channels}

\label{app:cw-as-qw}Suppose that $p\left(  y,z|x\right)  $ is a classical
wiretap channel such that $x$ is the input and $y$ and $z$ are the outputs for
the legitimate receiver and the wiretapper, respectively. Then we can embed
the random variables $X$, $Y$, and $Z$ into quantum systems, so that the
resulting wiretap channel has the following action on an arbitrary input state
$\rho$:%
\begin{equation}
\mathcal{N}_{\text{C}}^{A^{\prime}\rightarrow BE}\left(  \rho\right)
\equiv\sum_{x,y,z}\left\langle x\right\vert \rho\left\vert x\right\rangle
\ p\left(  y,z|x\right)  \left\vert y\right\rangle \left\langle y\right\vert
^{B}\otimes\left\vert z\right\rangle \left\langle z\right\vert ^{E}.
\label{eq:classical-wiretap}%
\end{equation}
The physical interpretation of the above channel is that it first
\textit{measures} the input system in the orthonormal basis $\left\{
\left\vert x\right\rangle \left\langle x\right\vert \right\}  $ (ensuring that
the input is effectively classical) and \textit{prepares} the classical states
$\left\vert y\right\rangle ^{B}$ and $\left\vert z\right\rangle ^{E}$ for Bob
and Eve with probability $p\left(  y,z|x\right)  $. One can check that the
Kraus operators \cite{W11}\ for this classical channel are%
\[
\left\{  \sqrt{p\left(  y,z|x\right)  }\left(  \left\vert y\right\rangle
^{B}\otimes\left\vert z\right\rangle ^{E}\right)  \left\langle x\right\vert
^{A^{\prime}}\right\}  _{x,y,z}.
\]
Thus, by a standard construction \cite{W11}, an isometric extension of this
classical wiretap channel acts as follows on a pure state input $\left\vert
\psi\right\rangle $:%
\begin{align*}
U_{\mathcal{N}_{\text{C}}}^{A^{\prime}\rightarrow BES_{2}}\left\vert
\psi\right\rangle =
\sum_{x,y,z}\sqrt{p\left(  y,z|x\right)  }\left(  \left\vert y\right\rangle
^{B}\otimes\left\vert z\right\rangle ^{E}\right)  \left\langle x\right\vert
^{A^{\prime}}\left\vert \psi\right\rangle \otimes\left\vert x,y,z\right\rangle
^{S_{2}},
\end{align*}
so that tracing over system $S_{2}$ recovers the action of the original
channel in (\ref{eq:classical-wiretap}).\footnote{The square root of a
probability might seem strange at first glance when appearing in an evolution,
but it is in fact quite natural in quantum information theory, being
interpreted physically as a probability amplitude.}

\section{Proof of Theorem~\ref{thm:suff-cond}}

\label{app:no-secret-key-proof}As described in
Section~\ref{sec:polarizaton-review}, one can prove that channel polarization
takes hold by considering the channel splitting and combining process as a
random birth process $\left\{  W_{n}:n\geq0\right\}  $ (with the channel
choice picked by some IID\ Bernoulli process $\left\{  B_{n}:n\geq1\right\}  $
and setting $W_{0}=W$). One can then consider the induced birth process%
\[
\left\{  F_{n}:n\geq0\right\}  \equiv\{\sqrt{F\left(  W_{n}\right)  }%
:n\geq0\}
\]
for the fidelity channel parameter. The inequalities (26-27) from
Ref.~\cite{WG11}\ (an extension of Arikan's inequalities~\cite{A09})
demonstrate that the following extremal process $\left\{  F_{n}^{\prime}%
:n\geq0\right\}  $ bounds the actual channel process $\left\{  F_{n}%
:n\geq0\right\}  $:%
\[
F_{n+1}^{\prime}=\left\{
\begin{array}
[c]{ccc}%
F_{n}^{\prime2} & \text{if} & B_{n}=0\\
2F_{n}^{\prime}-F_{n}^{\prime2} & \text{if} & B_{n}=1
\end{array}
\right.  ,
\]
a relation which can be written more symmetrically as%
\begin{align}
F_{n+1}^{\prime}  &  =F_{n}^{\prime2}\ \ \ \ \ \ \ \ \ \ \ \text{if\ \ \ }%
B_{n}=0,\nonumber\\
1-F_{n+1}^{\prime}  &  =\left(  1-F_{n}^{\prime}\right)  ^{2}%
\ \ \ \text{if\ \ \ }B_{n}=1. \label{eq:symmetric-form-birth-proc}%
\end{align}
From now on, we make abbreviations such as $\left\{  F_{n}\right\}  =\left\{
F_{n}:n\geq0\right\}  $ in order to simplify the notation.

The extremal process above has the nice property~\cite[Observation 4
(ii)]{AT09} (see the arXiv version) that for every realization $\left\{
b_{n}\right\}  $ of the process $\left\{  B_{n}\right\}  $ (and thus for every
realization $\left\{  f_{n}^{\prime}\right\}  $ of $\left\{  F_{n}^{\prime
}\right\}  $) there exists a particular initial threshold value $F_{\text{th}%
}^{\prime}\left(  \left\{  b_{n}\right\}  \right)  $ such that either%
\[
\lim_{n\rightarrow\infty}f_{n}^{\prime}=0\text{ if }F_{0}^{\prime
}<F_{\text{th}}^{\prime}\left(  \left\{  b_{n}\right\}  \right)  ,
\]
or%
\[
\lim_{n\rightarrow\infty}f_{n}^{\prime}=1\text{ if }F_{0}^{\prime}\geq
F_{\text{th}}^{\prime}\left(  \left\{  b_{n}\right\}  \right)  .
\]
(Note that $F_{0}^{\prime}$ is deterministic and is the initial value of the process.)

We can denote the respective fidelity processes for the amplitude and phase
channels in our coding scheme as $\left\{  F_{n}^{A}\right\}  $ and $\left\{
F_{n}^{P}\right\}  $ and the respective random birth processes as $\left\{
B_{n}^{A}\right\}  $ and $\left\{  B_{n}^{P}\right\}  $. Also, let $\left\{
F_{n}^{A\prime}\right\}  $ and $\left\{  F_{n}^{P\prime}\right\}  $ denote the
corresponding extremal processes. The important observation made in
Ref.~\cite{RDR11} is that the process $\left\{  F_{n}^{P}\right\}  $ makes the
opposite choice of channel at each step of the birth process because the phase
encoder is the reverse of the amplitude encoder. That is, it holds for every
$n$ and for every realization $\left\{  b_{n}^{A}\right\}  $ and $\left\{
b_{n}^{P}\right\}  $ that%
\[
b_{n}^{P}=1-b_{n}^{A}.
\]
Thus, we can write $B_{n}^{P}=1-B_{n}^{A}$, so that $B_{n}^{P}$ is completely
determined by $B_{n}^{A}$. The extremal amplitude channel process $\left\{
F_{n}^{A\prime}\right\}  $\ is already of the form in
(\ref{eq:symmetric-form-birth-proc}), and we can consider the extremal phase
process as $\left\{  1-F_{n}^{P\prime}\right\}  $ in order for it to have this
same form. Thus, a realization $\{f_{n}^{A^{\prime}}\}$ of the extremal
amplitude channel process $\left\{  F_{n}^{A\prime}\right\}  $ converges to
one if%
\[
F_{0}^{A\prime}\geq F_{\text{th}}^{\prime}(\{b_{n}^{A}\}),
\]
and a realization $\{1-f_{n}^{P^{\prime}}\}$ of the extremal phase process
$\left\{  1-F_{n}^{P\prime}\right\}  $ converges to zero if%
\[
1-F_{0}^{P\prime}<F_{\text{th}}^{\prime}(\{b_{n}^{A}\}),
\]
implying that $\left\{  f_{n}^{P\prime}\right\}  $ converges to one if%
\[
F_{0}^{P\prime}>1-F_{\text{th}}^{\prime}(\{b_{n}^{A}\}).
\]
Thus, the sum process $\left\{  F_{n}^{A\prime}+F_{n}^{P\prime}\right\}  $
converges to two if%
\begin{align}
F_{0}^{A\prime}+F_{0}^{P\prime} &  \geq F_{\text{th}}^{\prime}(\{b_{n}%
^{A}\})+1-F_{\text{th}}^{\prime}(\{b_{n}^{A}\})\nonumber\\
&  =1.\label{eq:condition-no-secret-bits}%
\end{align}
The above bound is a \textit{universal}, sufficient lower bound for the sum
process to converge to two, that holds regardless of the threshold value
$F_{\text{th}}^{\prime}(\{b_{n}^{A}\})$ for a particular realization
$\{b_{n}^{A}\}$. It follows that a given realization $\left\{  f_{n}^{A}%
+f_{n}^{P}\right\}  $ of the actual sum process $\left\{  F_{n}^{A}+F_{n}%
^{P}\right\}  $ can only converge to two when
(\ref{eq:condition-no-secret-bits}) holds because we set $F_{0}^{A^{\prime}%
}=F_{0}^{A}$ and the extremal process bounds the actual process (note that
some realizations might converge to one or zero as well). If a realization
$\left\{  f_{n}^{A}+f_{n}^{P}\right\}  $ of the sum process $\left\{
F_{n}^{A}+F_{n}^{P}\right\}  $ converges to two, then this implies that the
set $\mathcal{B}$ is non-empty, i.e., the code will require some secret key
bits. So, if the condition in the statement of the theorem holds, no
realization of the sum process can ever converge to two, and the code will not
require any secret key bits.

The above argument only holds in the asymptotic limit of many recursions of the encoding
such that the channel polarization effect takes hold (where all synthesized channels are
polarized to be completely perfect or useless). That is, the argument does not apply whenever there
is a finite number of recursions---in this case, if the number of recursions is large enough,
then a large fraction of synthesized channels polarize according to some tolerance, but there
is always a small fraction that have not polarized. Thus, we can only conclude
that the above proof applies in the limit of many recursions and that the rate of secret
key consumption vanishes in this limit. It is an open question to adapt the above argument
to the finite case, but we suspect that some form of it holds in this regime.

\section{Example Channels}

\label{app:example-channels}In this appendix, we evaluate the condition from Theorem~\ref{thm:suff-cond}%
\ for several examples, including the independent binary symmetric wiretap
channel model from Ref.~\cite{MV10}, the erasure wiretap channel, and the
amplitude damping channel. For many of these cases, we find that the resulting
polar codes have a vanishing secret key rate for a wide range of parameters.

\subsection{Binary Symmetric Wiretap Channel}

Suppose that the channel to Bob is a binary symmetric channel with flip
probability $p_{B}$ and the channel to Eve is an independent binary symmetric
channel with flip probability $p_{E}$. The symmetric private information for
this channel is equal to%
\[
\left(  1-H_{2}\left(  p_{B}\right)  \right)  -\left(  1-H_{2}\left(
p_{E}\right)  \right)  =H_{2}\left(  p_{E}\right)  -H_{2}\left(  p_{B}\right)
,
\]
which is only positive whenever $p_{B}<p_{E}$. The conditional distribution
$p\left(  y,z|x\right)  $ for this channel is just%
\begin{equation}
p\left(  y,z|x\right)  =\left(  p_{B}\right)  ^{x+y}\left(  1-p_{B}\right)
^{x+y+1}\left(  p_{E}\right)  ^{x+z}\left(  1-p_{E}\right)  ^{x+z+1},
\label{eq:BSC-wiretap}%
\end{equation}
where Alice inputs $x$, Bob receives output
$y$, Eve receives output $z$, and $y,z,x\in\left\{  0,1\right\}  $. Observe that the above channel factorizes as%
\[
p\left(  y,z|x\right)  =p\left(  y|x\right)  p\left(  z|x\right)  ,
\]
where%
\begin{align*}
p\left(  y|x\right)   &  =\left(  p_{B}\right)  ^{x+y}\left(  1-p_{B}\right)
^{x+y+1},\\
p\left(  z|x\right)   &  =\left(  p_{E}\right)  ^{x+z}\left(  1-p_{E}\right)
^{x+z+1}.
\end{align*}
Since the amplitude channel to Bob in this case is just a binary symmetric
channel with flip probability $p_{B}$, the root fidelity $\sqrt{F(W_{A,B})}$
just reduces to the classical Bhattacharya parameter:%
\[
\sqrt{F(W_{A,B})}=2\sqrt{p_{B}\left(  1-p_{B}\right)  }.
\]

We now need to compute the fidelity for the phase channel to Bob. Since Alice
inputs classical states $\left\vert z\right\rangle $ to the amplitude channel,
the states $\left\vert \psi_{z}\right\rangle $ in (\ref{eq:phase-input}) are
just equal to $\left\vert z\right\rangle $ (there is no shield system $S_{1}%
$). This implies that the quantum input for the phase channel is of the form%
\[
Z^{x^{\prime}}\frac{1}{\sqrt{2}}\sum_{z^{\prime}}\left\vert z^{\prime
}\right\rangle ^{C}\left\vert z^{\prime}\right\rangle ^{A^{\prime}}=\frac
{1}{\sqrt{2}}\sum_{z^{\prime}}\left(  -1\right)  ^{x^{\prime}\cdot z^{\prime}%
}\left\vert z^{\prime}\right\rangle ^{C}\left\vert z^{\prime}\right\rangle
^{A^{\prime}},
\]
if $x^{\prime}$ is the input bit. The isometric extension of this classical
wiretap channel is then just%
\[
\sum_{x,y,z}\sqrt{p\left(  y,z|x\right)  }\left(  \left\vert y\right\rangle
^{B}\otimes\left\vert z\right\rangle ^{E}\right)  \left\langle x\right\vert
^{A^{\prime}}\otimes\left\vert x,y,z\right\rangle ^{S_{2}},
\]
with $p\left(  y,z|x\right)  $ given by (\ref{eq:BSC-wiretap}), so that the
action on the above input state is%
\begin{align*}
x^{\prime}  &  \rightarrow\frac{1}{\sqrt{2}}\sum_{x,y,z,z^{\prime}}\left(
-1\right)  ^{x^{\prime}\cdot z^{\prime}}\sqrt{p\left(  y,z|x\right)
}\left\vert y\right\rangle ^{B}\left\vert z\right\rangle ^{E}\left\vert
z^{\prime}\right\rangle ^{C}\left\langle x|z^{\prime}\right\rangle
^{A^{\prime}}\otimes\left\vert x,y,z\right\rangle ^{S_{2}}\\
&  =\frac{1}{\sqrt{2}}\sum_{x,y,z}\left(  -1\right)  ^{x^{\prime}\cdot x}%
\sqrt{p\left(  y,z|x\right)  }\left\vert y\right\rangle ^{B}\left\vert
z\right\rangle ^{E}\left\vert x\right\rangle ^{C}\otimes\left\vert
x,y,z\right\rangle ^{S_{2}}.
\end{align*}
To simplify things, we will write $p\left(  x\right)  =1/2$, giving%
\[
x^{\prime}\rightarrow\sum_{x,y,z}\left(  -1\right)  ^{x^{\prime}\cdot x}%
\sqrt{p\left(  y,z|x\right)  p\left(  x\right)  }\left\vert y\right\rangle
^{B}\left\vert z\right\rangle ^{E}\left\vert x\right\rangle ^{C}%
\otimes\left\vert x,y,z\right\rangle ^{S_{2}}%
\]
Tracing over the $E$ system then gives the phase channel to Bob:%
\begin{align*}
&  \text{Tr}_{E}\left\{  \sum_{x,y,z,x^{\prime\prime},y^{\prime\prime
},z^{\prime\prime}}\left(  -1\right)  ^{x^{\prime}\cdot x}\left(  -1\right)
^{x^{\prime}\cdot x^{\prime\prime}}\sqrt{p\left(  y,z|x\right)  p\left(
x\right)  p\left(  y^{\prime\prime},z^{\prime\prime}|x^{\prime\prime}\right)
p\left(  x^{\prime\prime}\right)  }\left\vert y\right\rangle \left\langle
y^{\prime\prime}\right\vert ^{B}\left\vert z\right\rangle \left\langle
z^{\prime\prime}\right\vert ^{E}\left\vert x\right\rangle \left\langle
x^{\prime\prime}\right\vert ^{C}\otimes\left\vert x,y,z\right\rangle
\left\langle x^{\prime\prime},y^{\prime\prime},z^{\prime\prime}\right\vert
^{S_{2}}\right\} \\
&  =\sum_{x,y,z,x^{\prime\prime},y^{\prime\prime}}\left(  -1\right)
^{x^{\prime}\cdot\left(  x+x^{\prime\prime}\right)  }\sqrt{p\left(
y,z|x\right)  p\left(  x\right)  p\left(  y^{\prime\prime},z|x^{\prime\prime
}\right)  p\left(  x^{\prime\prime}\right)  }\left\vert y\right\rangle
\left\langle y^{\prime\prime}\right\vert ^{B}\otimes\left\vert x\right\rangle
\left\langle x^{\prime\prime}\right\vert ^{C}\otimes\left\vert
x,y,z\right\rangle \left\langle x^{\prime\prime},y^{\prime\prime},z\right\vert
^{S_{2}}\\
&  =\sum_{x,y,z,x^{\prime\prime},y^{\prime\prime}}\left(  -1\right)
^{x^{\prime}\cdot\left(  x+x^{\prime\prime}\right)  }\sqrt{p\left(
y|x\right)  p\left(  z|x\right)  p\left(  x\right)  p\left(  y^{\prime\prime
}|x^{\prime\prime}\right)  p\left(  z|x^{\prime\prime}\right)  p\left(
x^{\prime\prime}\right)  }\left\vert y\right\rangle \left\langle
y^{\prime\prime}\right\vert ^{B}\otimes\left\vert x\right\rangle \left\langle
x^{\prime\prime}\right\vert ^{C}\otimes\left\vert x,y,z\right\rangle
\left\langle x^{\prime\prime},y^{\prime\prime},z\right\vert ^{S_{2}}\\
&  =\sum_{x,y,z,x^{\prime\prime},y^{\prime\prime}}\left(  -1\right)
^{x^{\prime}\cdot\left(  x+x^{\prime\prime}\right)  }\sqrt{p\left(
y|x\right)  p\left(  x|z\right)  p\left(  z\right)  p\left(  y^{\prime\prime
}|x^{\prime\prime}\right)  p\left(  x^{\prime\prime}|z\right)  p\left(
z\right)  }\left\vert y\right\rangle \left\langle y^{\prime\prime}\right\vert
^{B}\otimes\left\vert x\right\rangle \left\langle x^{\prime\prime}\right\vert
^{C}\otimes\left\vert x,y,z\right\rangle \left\langle x^{\prime\prime
},y^{\prime\prime},z\right\vert ^{S_{2}}%
\end{align*}
We can then factorize the above state as follows:%
\[
\omega_{x^{\prime}}^{BCS_{X}S_{Y}S_{Z}}\equiv\sum_{z}p\left(  z\right)
\left\vert \chi_{z,x^{\prime}}\right\rangle \left\langle \chi_{z,x^{\prime}%
}\right\vert ^{BCS_{X}S_{Y}}\otimes\left\vert z\right\rangle \left\langle
z\right\vert ^{S_{Z}},
\]
where%
\[
\left\vert \chi_{z,x^{\prime}}\right\rangle ^{BCS_{X}S_{Y}}\equiv\frac
{1}{\sqrt{2}}\sum_{x,y}\left(  -1\right)  ^{x^{\prime}\cdot x}\sqrt{p\left(
y|x\right)  p\left(  x|z\right)  }\left\vert y\right\rangle ^{B}\left\vert
x\right\rangle ^{C}\left\vert x\right\rangle ^{S_{X}}\left\vert y\right\rangle
^{S_{Y}}.
\]
Noting that $p\left(  z\right)  =1/2$, we now compute the fidelity
$\sqrt{F(W_{P,B})}$ as%
\begin{align*}
\sqrt{F(W_{P,B})}  &  =\sqrt{F\left(  \omega_{0}^{BCS_{X}S_{Y}S_{Z}}%
,\omega_{1}^{BCS_{X}S_{Y}S_{Z}}\right)  }\\
&  =\sum_{z}\frac{1}{2}\sqrt{F\left(  \chi_{z,0},\chi_{z,1}\right)  }\\
&  =\sum_{z}\frac{1}{2}\left\vert \left\langle \chi_{z,0}|\chi_{z,1}%
\right\rangle \right\vert \\
&  =\left\vert p_{E}-1/2\right\vert ,
\end{align*}
where the last line follows from%
\begin{align*}
\left\vert \left\langle \chi_{z,0}|\chi_{z,1}\right\rangle \right\vert  &
=\frac{1}{2}\left\vert \sum_{x,y,x^{\prime\prime},y^{\prime\prime}}\left(
-1\right)  ^{x}\sqrt{p\left(  y^{\prime\prime}|x^{\prime\prime}\right)
p\left(  x^{\prime\prime}|z\right)  p\left(  y|x\right)  p\left(  x|z\right)
}\left\langle y^{\prime\prime}\right\vert \left\vert y\right\rangle
^{B}\left\langle x^{\prime\prime}\right\vert \left\vert x\right\rangle
^{C}\left\langle x^{\prime\prime}\right\vert \left\vert x\right\rangle
^{S_{X}}\left\langle y^{\prime\prime}\right\vert \left\vert y\right\rangle
^{S_{Y}}\right\vert \\
&  =\frac{1}{2}\left\vert \sum_{x,y}\left(  -1\right)  ^{x}p\left(
y|x\right)  p\left(  x|z\right)  \right\vert \\
&  =\frac{1}{2}\left\vert \sum_{x}\left(  -1\right)  ^{x}p\left(  x|z\right)
\right\vert \\
&  =\frac{1}{2}\left\vert \sum_{x}\left(  -1\right)  ^{x}\left(  p_{E}\right)
^{x+z}\left(  1-p_{E}\right)  ^{x+z+1}\right\vert \\
&  =\left\vert p_{E}-1/2\right\vert .
\end{align*}
The relation $p\left(  x|z\right)  =p\left(  z|x\right)  $ follows because
$p\left(  x\right)  =p\left(  z\right)  =1/2$.

Thus, from Theorem~\ref{thm:suff-cond}, the sufficient condition for the
wiretap channel in (\ref{eq:BSC-wiretap}) to have a vanishing secret key rate is
just%
\[
2\sqrt{p_{B}\left(  1-p_{B}\right)  }+\left\vert p_{E}-1/2\right\vert <1.
\]
If we assume that $p_{E}<1/2$, then this reduces to%
\[
2\sqrt{p_{B}\left(  1-p_{B}\right)  }<1/2+p_{E}.
\]

\begin{figure}
[ptb]
\begin{center}
\includegraphics[
natheight=3.440200in,
natwidth=5.040100in,
height=2.3912in,
width=3.4904in
]%
{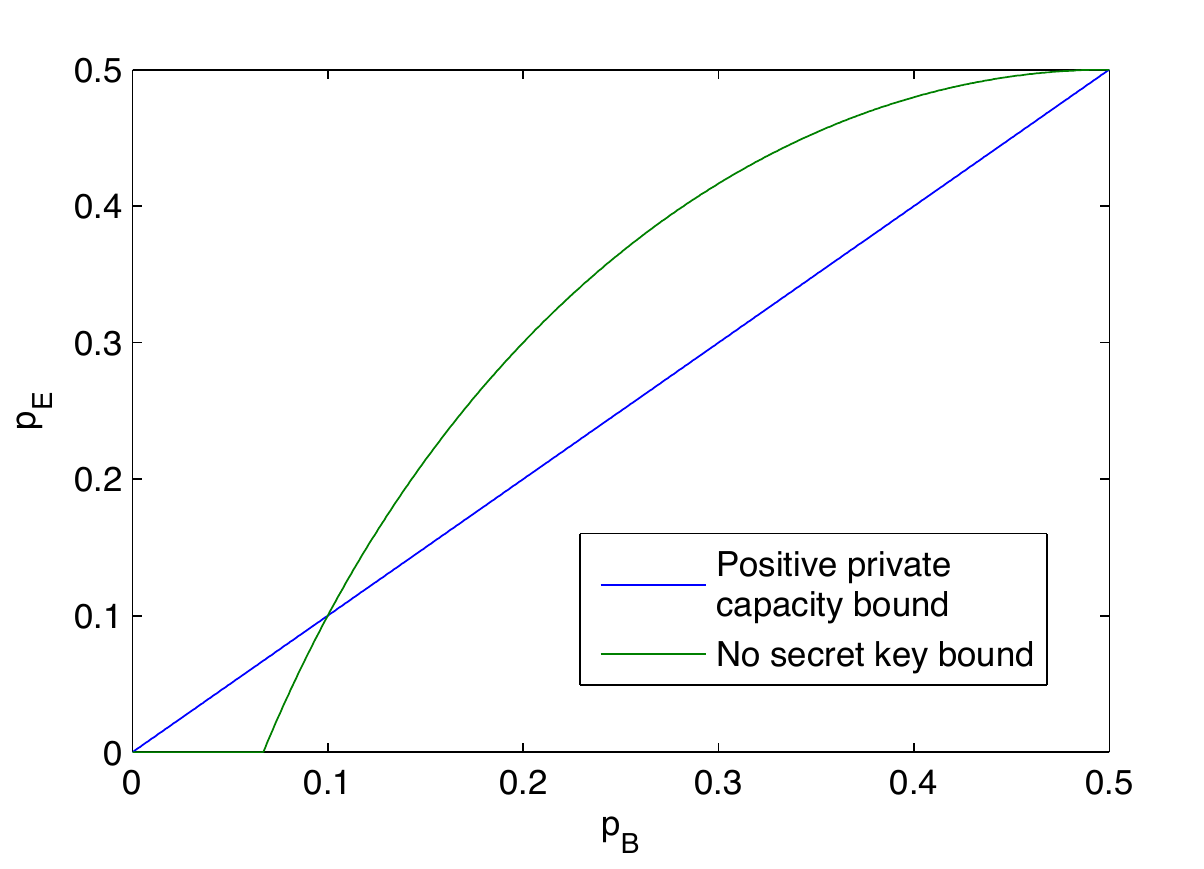}%
\caption{The parameters $p_B$ and $p_E$ correspond to bit flip probabilities
in the binary symmetric wiretap channel. Below the blue line is the region where the symmetric private
information is positive. Below the green line is the region for when the secret key rate of the code
vanishes according to Theorem~\ref{thm:suff-cond}. The
result is that the polar code has a vanishing secret key rate for a wide range of
interesting parameters $p_B$ and $p_E$.}%
\label{fig:BSC-wiretap}%
\end{center}
\end{figure}

Figure~\ref{fig:BSC-wiretap}\ provides a plot of two bounds:\ a bound for when
the symmetric private information is positive and bound for when the polar
code has a vanishing secret key rate. We find that the secret key rate vanishes for a
wide range of parameters.%

\subsection{Erasure Channel}

An erasure channel transmits the input bit with probability $1-\epsilon$ and
provides an erasure symbol $e$ to the receiver (different from 0 or 1) with
probability $\epsilon$. A simple model for a channel to the wiretapper is to
give the input bit to the wiretapper whenever the receiver gets the erasure
symbol (thus, the channel to the wiretapper is an erasure channel with erasure
probability $1-\epsilon$).

It turns out that this model is equivalent to the quantum erasure channel:%
\[
\rho\rightarrow\left(  1-\epsilon\right)  \rho+\epsilon\left\vert
e\right\rangle \left\langle e\right\vert ,
\]
for which it is well known that the complementary channel is just%
\[
\rho\rightarrow\epsilon\rho+\left(  1-\epsilon\right)  \left\vert
e\right\rangle \left\langle e\right\vert ,
\]
so that this channel is equivalent to the wiretap channel mentioned above.
Both the quantum and private capacities of this channel are equal to
$1-2\epsilon$.

The amplitude channel to Bob is just an erasure channel with erasure
probability $\epsilon$. We now consider the phase channel to Bob:%
\[
x\rightarrow\mathcal{N}\left(  Z^{x}\left\vert \Phi\right\rangle \left\langle
\Phi\right\vert Z^{x}\right)  =\epsilon Z^{x}\left\vert \Phi\right\rangle
\left\langle \Phi\right\vert Z^{x}+\left(  1-\epsilon\right)  \pi
\otimes\left\vert e\right\rangle \left\langle e\right\vert ,
\]
where $\pi$ is the maximally mixed state. It is clear that the above channel
is also just an erasure channel with erasure probability $\epsilon$ because
Bob can perform the measurement $\left\{  \left\vert 0\right\rangle
\left\langle 0\right\vert +\left\vert 1\right\rangle \left\langle 1\right\vert
,\left\vert e\right\rangle \left\langle e\right\vert \right\}  $ on the second
qubit to determine if he received the state. If he does receive it, he can
then perform a Bell measurement to retrieve the bit $x$.

Since the root fidelity of an erasure channel is just $\epsilon$, the condition
from Theorem~\ref{thm:suff-cond} for vanishing secret key rate just reduces to%
\[
2\epsilon<1,
\]
which is satisfied for all erasure probabilities $\epsilon$ for which the
private information is positive.

\subsection{Amplitude Damping Channel}

The last example of a channel that we study is the amplitude damping channel.
This channel models photon loss when transmitting a state in the zero or
single-photon subspace over a pure-loss bosonic channel (a beamsplitter with transmissivity
$\eta$). The Kraus operators for this channel are%
\begin{align*}
A_{0}  & \equiv\sqrt{1-\eta}\left\vert 0\right\rangle \left\langle
1\right\vert ,\\
A_{1}  & \equiv\left\vert 0\right\rangle \left\langle 0\right\vert +\sqrt
{\eta}\left\vert 1\right\rangle \left\langle 1\right\vert ,
\end{align*}
so that the evolution of an input qubit is%
\[
\rho\rightarrow A_{0}\rho A_{0}^{\dag}+A_{1}\rho A_{1}^{\dag}.
\]
This channel has a positive private capacity whenever $\eta>1/2$. If $\eta
\leq1/2$, then the majority of the output is going to the wiretapper (or the
environment of the channel) and so there is not any positive private
capacity for this parameter range.

The amplitude channel to Bob is%
\begin{align*}
0  & \rightarrow A_{0}\left\vert 0\right\rangle \left\langle 0\right\vert
A_{0}^{\dag}+A_{1}\left\vert 0\right\rangle \left\langle 0\right\vert
A_{1}^{\dag}=\left\vert 0\right\rangle \left\langle 0\right\vert ,\\
1  & \rightarrow A_{0}\left\vert 1\right\rangle \left\langle 1\right\vert
A_{0}^{\dag}+A_{1}\left\vert 1\right\rangle \left\langle 1\right\vert
A_{1}^{\dag}=\left(  1-\eta\right)  \left\vert 0\right\rangle \left\langle
0\right\vert +\eta\left\vert 1\right\rangle \left\langle 1\right\vert .
\end{align*}
Thus, the fidelity for this amplitude channel is $\sqrt{1-\eta}$.

Now consider the phase channel:%
\[
x\rightarrow\mathcal{N}\left(  Z^{x}\left\vert \Phi\right\rangle \left\langle
\Phi\right\vert Z^{x}\right)  ,
\]
where $\left\vert \Phi\right\rangle $ is the Bell state. Analyzing for this
case gives%
\[
\left(  1-\eta\right)  \left\vert 0\right\rangle \left\langle 1\right\vert
\left(  Z^{x}\left\vert \Phi\right\rangle \left\langle \Phi\right\vert
Z^{x}\right)  \left\vert 1\right\rangle \left\langle 0\right\vert +\left(
\left\vert 0\right\rangle \left\langle 0\right\vert +\sqrt{1-\gamma}\left\vert
1\right\rangle \left\langle 1\right\vert \right)  \left(  Z^{x}\left\vert
\Phi\right\rangle \left\langle \Phi\right\vert Z^{x}\right)  \left(
\left\vert 0\right\rangle \left\langle 0\right\vert +\sqrt{1-\gamma}\left\vert
1\right\rangle \left\langle 1\right\vert \right)
\]
We handle the first term:%
\begin{align*}
\left(  1-\eta\right)  \left\vert 0\right\rangle \left\langle 1\right\vert
\left(  Z^{x}\left\vert \Phi\right\rangle \left\langle \Phi\right\vert
Z^{x}\right)  \left\vert 1\right\rangle \left\langle 0\right\vert  &
=\frac{1-\eta}{2}\left\vert 0\right\rangle \left\langle 1\right\vert \left(
\left\vert 00\right\rangle \left\langle 00\right\vert +\left(  -1\right)
^{x}\left\vert 11\right\rangle \left\langle 00\right\vert +\left(  -1\right)
^{x}\left\vert 00\right\rangle \left\langle 11\right\vert +\left\vert
11\right\rangle \left\langle 11\right\vert \right)  \left\vert 1\right\rangle
\left\langle 0\right\vert \\
&  =\frac{1-\eta}{2}\left\vert 10\right\rangle \left\langle 10\right\vert
\end{align*}
We handle the second term:%
\begin{align*}
&  \left(  \left\vert 0\right\rangle \left\langle 0\right\vert +\sqrt{\eta
}\left\vert 1\right\rangle \left\langle 1\right\vert \right)  \left(
Z^{x}\left\vert \Phi\right\rangle \left\langle \Phi\right\vert Z^{x}\right)
\left(  \left\vert 0\right\rangle \left\langle 0\right\vert +\sqrt{\eta
}\left\vert 1\right\rangle \left\langle 1\right\vert \right)  \\
&  =\frac{1}{2}\left(  \left\vert 0\right\rangle \left\langle 0\right\vert
+\sqrt{\eta}\left\vert 1\right\rangle \left\langle 1\right\vert \right)
\left(  \left\vert 00\right\rangle \left\langle 00\right\vert +\left(
-1\right)  ^{x}\left\vert 11\right\rangle \left\langle 00\right\vert +\left(
-1\right)  ^{x}\left\vert 00\right\rangle \left\langle 11\right\vert
+\left\vert 11\right\rangle \left\langle 11\right\vert \right)  \left(
\left\vert 0\right\rangle \left\langle 0\right\vert +\sqrt{\eta}\left\vert
1\right\rangle \left\langle 1\right\vert \right)  \\
&  =\frac{1}{2}\left(  \left\vert 00\right\rangle \left\langle 00\right\vert
+\left(  -1\right)  ^{x}\sqrt{\eta}\left\vert 11\right\rangle \left\langle
00\right\vert +\left(  -1\right)  ^{x}\left\vert 00\right\rangle \left\langle
11\right\vert +\sqrt{\eta}\left\vert 11\right\rangle \left\langle
11\right\vert \right)  \left(  \left\vert 0\right\rangle \left\langle
0\right\vert +\sqrt{\eta}\left\vert 1\right\rangle \left\langle 1\right\vert
\right)  \\
&  =\frac{1}{2}\left(  \left\vert 00\right\rangle \left\langle 00\right\vert
+\left(  -1\right)  ^{x}\sqrt{\eta}\left\vert 11\right\rangle \left\langle
00\right\vert +\left(  -1\right)  ^{x}\sqrt{\eta}\left\vert 00\right\rangle
\left\langle 11\right\vert +\eta\left\vert 11\right\rangle \left\langle
11\right\vert \right)
\end{align*}
Adding the two terms gives%
\[
\frac{1}{2}\left(  \left\vert 00\right\rangle \left\langle 00\right\vert
+\left(  -1\right)  ^{x}\sqrt{\eta}\left\vert 11\right\rangle \left\langle
00\right\vert +\left(  -1\right)  ^{x}\sqrt{\eta}\left\vert 00\right\rangle
\left\langle 11\right\vert +\eta\left\vert 11\right\rangle \left\langle
11\right\vert \right)  +\frac{1-\eta}{2}\left\vert 10\right\rangle
\left\langle 10\right\vert
\]
which has the following matrix representation in the computational basis:%
\[%
\begin{bmatrix}
\frac{1}{2} & 0 & 0 & \frac{1}{2}\left(  -1\right)  ^{x}\sqrt{\eta}\\
0 & 0 & 0 & 0\\
0 & 0 & \frac{1-\eta}{2} & 0\\
\frac{1}{2}\left(  -1\right)  ^{x}\sqrt{\eta} & 0 & 0 & \frac{\eta}{2}%
\end{bmatrix}
.
\]

\begin{figure}
[ptb]
\begin{center}
\includegraphics[
natheight=4.373400in,
natwidth=5.833200in,
height=2.6161in,
width=3.4809in
]%
{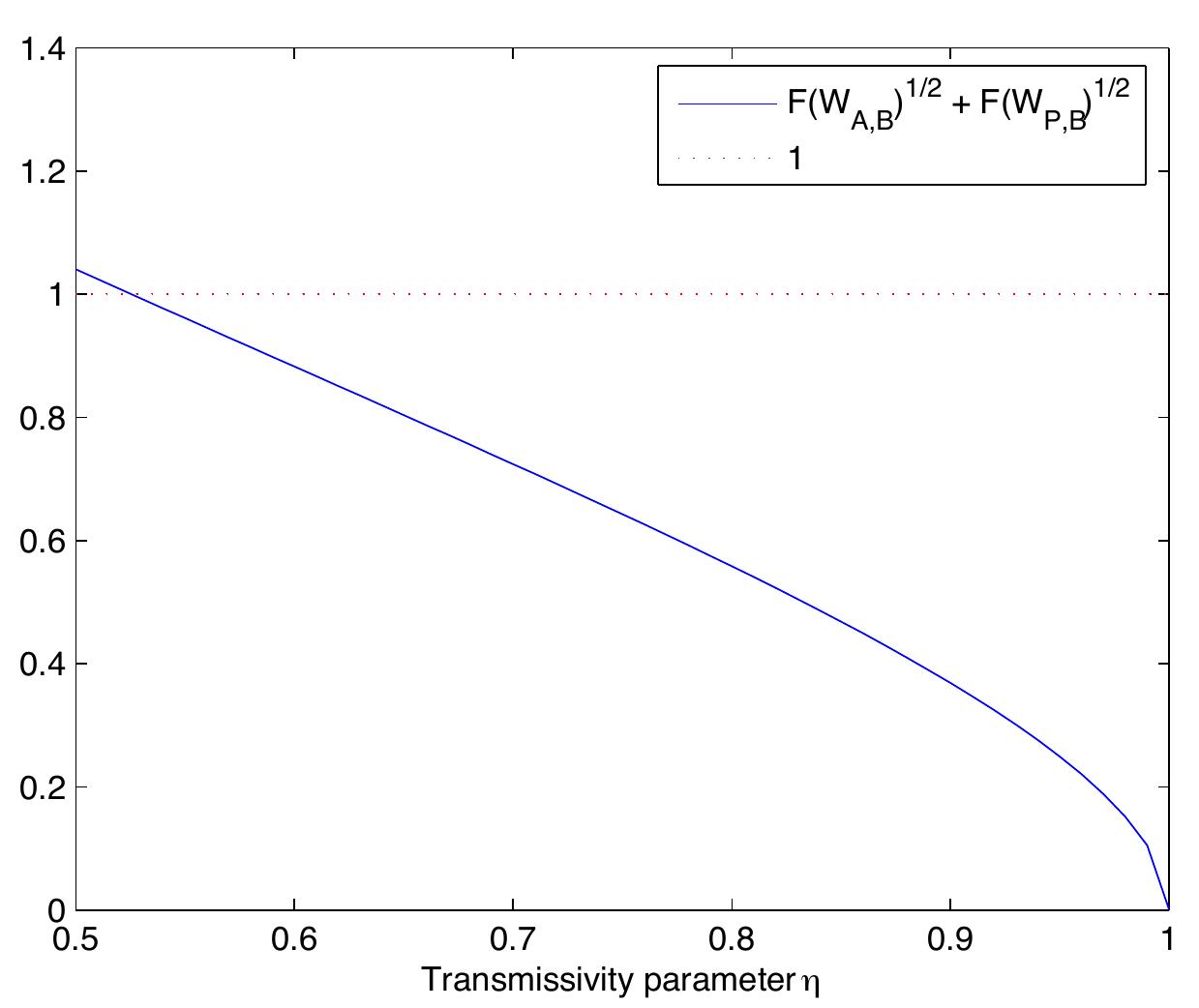}%
\caption{A comparison of the quantity $\sqrt{F(W_{A,B})}+\sqrt{F(W_{P,B})}$ from Theorem~\ref{thm:suff-cond}
and 1 for various transmissivities $\eta$ of the amplitude damping channel. The result is that a polar code
has a vanishing secret key rate for most $\eta$.}
\label{fig:amp-channel-no-secret-key}%
\end{center}
\end{figure}

We numerically compute the fidelity for the phase channel, and the plot in
Figure~\ref{fig:amp-channel-no-secret-key}\ shows all of the damping
parameters which meet the condition from Theorem~\ref{thm:suff-cond}. The
result is that the polar has a vanishing secret key rate for most transmissivities~$\eta$.

\end{document}